\documentclass[conference]{IEEEtran}
\IEEEoverridecommandlockouts
\usepackage{cite}
\usepackage{amsmath,amssymb,amsfonts}
\usepackage{algorithmic}
\usepackage{graphicx}
\usepackage{textcomp}
\usepackage{hyperref}
\usepackage{xcolor}
\def\BibTeX{{\rm B\kern-.05em{\sc i\kern-.025em b}\kern-.08em
    T\kern-.1667em\lower.7ex\hbox{E}\kern-.125emX}}

\usepackage{listings}
\definecolor{codegreen}{rgb}{0,0.6,0}
\definecolor{codegray}{rgb}{0.5,0.5,0.5}
\definecolor{codepurple}{rgb}{0.58,0,0.82}
\definecolor{backcolour}{rgb}{0.95,0.95,0.95}

\definecolor{delim}{RGB}{20,105,176}
\definecolor{numb}{RGB}{106, 109, 32}
\definecolor{string}{rgb}{0.64,0.08,0.08}

\lstdefinestyle{qfaasstyle}{
    backgroundcolor=\color{backcolour},   
    commentstyle=\color{codegreen},
    keywordstyle=\color{magenta},
    numberstyle=\tiny\color{codegray},
    stringstyle=\color{codepurple},
    basicstyle=\ttfamily\footnotesize,
    breakatwhitespace=false,         
    breaklines=true,                 
    captionpos=b,
    title={},
    keepspaces=true,                 
    numbers=none,                    
    numbersep=10pt,                  
    showspaces=false,                
    showstringspaces=false,
    showtabs=false,                  
    tabsize=1,
    frame=single,
    xleftmargin=0.1in,
    xrightmargin=0.1in
}

\lstdefinelanguage{java}{
    rulecolor=\color{black},
    postbreak=\raisebox{0ex}[0ex][0ex]{\ensuremath{\color{gray}\hookrightarrow\space}},
    upquote=true,
    morestring=[b]",
    literate=
     *{0}{{{\color{numb}0}}}{1}
      {1}{{{\color{numb}1}}}{1}
      {2}{{{\color{numb}2}}}{1}
      {3}{{{\color{numb}3}}}{1}
      {4}{{{\color{numb}4}}}{1}
      {5}{{{\color{numb}5}}}{1}
      {6}{{{\color{numb}6}}}{1}
      {7}{{{\color{numb}7}}}{1}
      {8}{{{\color{numb}8}}}{1}
      {9}{{{\color{numb}9}}}{1}
      {\{}{{{\color{delim}{\{}}}}{1}
      {\}}{{{\color{delim}{\}}}}}{1}
      {[}{{{\color{delim}{[}}}}{1}
      {]}{{{\color{delim}{]}}}}{1},
}

\newcommand*{\code}{\lstinline[keywordstyle=\color{blue}, basicstyle=\ttfamily\small\color{black}]}

\begin{document}

\title{iQuantum: A Case for Modeling and Simulation of Quantum Computing Environments
}

\author{
    \IEEEauthorblockN{Hoa T. Nguyen\textsuperscript{1}, Muhammad Usman\textsuperscript{2,3}, Rajkumar Buyya\textsuperscript{1}}
    \IEEEauthorblockA{\textsuperscript{1}Cloud Computing and Distributed Systems (CLOUDS) Laboratory, School of Computing and Information Systems,\\ The University of Melbourne, Parkville, 3052, Victoria, Australia}
    \IEEEauthorblockA{\textsuperscript{2}School of Physics, The University of Melbourne, Parkville, 3052, Victoria, Australia}
    \IEEEauthorblockA{\textsuperscript{3}Data61, CSIRO, Clayton, 3168, Victoria, Australia}
}


\maketitle

\begin{abstract}
Today's quantum computers are primarily accessible through the cloud and potentially shifting to the edge network in the future. With the rapid advancement and proliferation of quantum computing research worldwide, there has been a considerable increase in demand for using cloud-based quantum computation resources. This demand has highlighted the need for designing efficient and adaptable resource management strategies and service models for quantum computing. However, the limited quantity, quality, and accessibility of quantum resources pose significant challenges to practical research in quantum  software and systems. To address these challenges, we propose iQuantum, a first-of-its-kind simulation toolkit that can model hybrid quantum-classical computing environments for prototyping and evaluating system design and scheduling algorithms. This paper presents the quantum computing system model, architectural design, proof-of-concept implementation, potential use cases, and future development of iQuantum. Our proposed iQuantum simulator is anticipated to boost research in quantum software and systems, particularly in the creation and evaluation of policies and algorithms for resource management, job scheduling, and hybrid quantum-classical task orchestration in quantum computing environments integrating edge and cloud resources.

\end{abstract}


\section{Introduction}

Quantum computing is an emerging field with tremendous potential to solve computationally intractable problems that can take years to solve with classical supercomputers. Its applications can revolutionize many fields, such as drug discovery \cite{zinner_quantum_2021}, finance \cite{griffin_quantum_2021}, optimization \cite{moll_quantum_2018}, and machine learning \cite{debenedictis_future_2018}. However, quantum computing is still under development, and there are numerous challenges in realizing its potential to bring its advantages to industry and the practical world \cite{quantum_technology_and_application_consortium__qutac_industry_2021}. Furthermore, one of the primary challenges of quantum computing is its difficulty in operation, requiring extreme environmental conditions \cite{krantz_quantum_2019}, specialized expertise, and massive investments in quantum hardware. 

Fortunately, the emergence and development of the cloud-based quantum computing (or quantum cloud) model provide a viable solution to make quantum computing accessible to the general public \cite{devitt_performing_2016}. Quantum cloud is the intersection of quantum computing and cloud computing, allowing access to quantum computation resources over the cloud without significant upfront investment in quantum hardware \cite{leymann_quantum_2020}. Many giant cloud providers, such as Microsoft Azure Quantum, IBM Quantum, and Amazon Braket, have started offering quantum computing services. In light of this quantum service paradigm shift, we can anticipate significant progress in the field of quantum software engineering in the near future.

As the demand for quantum computing services has risen rapidly, efficient cloud systems and resource management strategies are becoming increasingly necessary for optimal quantum resource utilization \cite{ravi_quantum_2021}. While physical quantum hardware can be accessed through the cloud, its high cost and limited availability pose challenges for large-scale evaluation of resource management strategy and can lead to insufficient experimental validation. Therefore, developing a simulation framework to model hybrid quantum computing environments is crucial to support research in cloud-based quantum software and systems. However, existing quantum simulators primarily focus on simulating the physical operation of the quantum system \cite{altman_quantum_2021} (quantum operation) rather than modeling quantum systems and services in quantum computing environments. As a result, research in system design and resource management for quantum cloud computing remains challenging. Without quantum computing simulation and modeling tools, researchers may find testing their system design or job scheduling algorithms difficult or even unattainable in a realistic quantum computing environment. Besides, without the standard simulator, it can be complicated to reproduce experimental results or compare the performance of different algorithms or applications consistently and meaningfully.

To address these challenges, we propose \textit{iQuantum}, a simulation framework for modeling quantum computing environments and facilitating resource management and system design. We propose the quantum system model and its resource management problems, which foster the development of a quantum modeling simulator. iQuantum has the potential to enable researchers to test and validate their algorithms and applications in a simulated quantum computing environment without the need for spending expensive costs to use quantum hardware. Besides, it can empower research and experiment in quantum software engineering, making it easier to compare results and reproduce experiments for more rigorous and impactful studies aligned with the latest advances in quantum computing. This study aims to address two research questions:
\begin{itemize}
    \item \textbf{RQ1. }\textit{Can we model a hybrid quantum-classical system in a cloud computing environment similar to modeling a classical paradigm?}
    
    RQ1 aims to investigate the feasibility of modeling today's cloud-based quantum computing environments. This is the key question we have thoroughly studied before embarking on the prototyping and implementation of the simulator. After realizing the possibility of addressing this question, we continue to identify \textit{what are the key elements of hybrid quantum system} and \textit{how to model them}. The detailed results of our studies on these questions are presented in Section \ref{sec:modeling-qcloud} and \ref{sec:model-design}.
    \item \textbf{RQ2.} \textit{Can we use the proposed simulator to model any use cases in practice}?
    
    RQ2 aims to identify potential use cases for the proposed simulator to consider its usefulness in quantum computing research. Our investigation to address this question is outlined in Section \ref{sec:model-design}.

\end{itemize}

By addressing these research questions, we aim to provide viable insights into the potential of designing and developing the proposed simulator for hybrid quantum computing. The main contributions and novelty of our study are as follows:
\begin{itemize}
    \item We derived the first conceptual model for quantum computing environments using the various metrics of today's quantum computers and quantum job execution on the cloud. This model provides a theoretical basis for simulating and analyzing quantum computing systems, which has not been previously formulated in the literature.
    
    \item We proposed the architecture design and proof-of-concept implementation of core components in iQuantum based on CloudSim \cite{calheiros_cloudsim_2011}, which is a well-known simulator for classical cloud system modeling. 
    
    \item We present various potential applications of iQuantum for modeling cloud-based quantum computing environments, which support research in system design and resource management, including quantum job scheduling, qubit mapping, and task orchestration in a hybrid quantum-classical cloud. These applications demonstrate the practical usefulness of our proposal for researchers and practitioners in the field.
\end{itemize}

The rest of the paper is organized as follows: In Section II, we introduce the related background of our study, including the concept of cloud-based quantum computing, the QCaaS model, quantum applications, and performance metrics used to benchmark present quantum computers. In Section III, we propose the system model of a quantum computing environment. Then, we suggest the system design and core implementation of iQuantum to validate the proposed quantum system model in Section IV. In Section V, we illustrate the use of iQuantum simulator in creating a model and simulation of a quantum node and quantum tasks. We discuss potential use cases and extensions of iQuantum, including job scheduling, qubit mapping, and modeling hybrid quantum-classical cloud systems in Section VI. We review other related work in the literature in Section VII and conclude our study with future work in Section VIII.

\section{Background}
\subsection{Hybrid Quantum Computing and QCaaS Models}
Cloud-based Quantum Computing (quantum cloud) is a paradigm in which quantum computation is performed on a quantum backend, such as a quantum computer or quantum simulator, over the cloud. These backends can be publicly accessible over the internet (public quantum cloud) or exclusively accessible within a specific organization using a privileged network (private quantum cloud). QCC provides a means for users to execute quantum tasks without needing to manage their own quantum hardware \cite{leymann_quantum_2020}. Quantum cloud is currently the only viable way to access today's quantum computers, due to the stringent requirements \cite{krantz_quantum_2019} for operating a quantum system to avoid environmental interactions that can affect quantum execution. In an ideal scenario, the quantum cloud can be fully accommodated by a quantum internet \cite{singh_quantum_internet_2021}, where all quantum data communications are processed through the quantum-based network without dependence on classical computers. However, the quantum internet is still in its early stage of development, with numerous challenges \cite{wehner_quantum_2018}. As a result, today's cloud-based quantum computers still rely on the classical internet and classical computers, which can be categorized as a hybrid quantum-classical cloud. 

\begin{figure}[htbp]
\centerline{\includegraphics[scale=0.07]{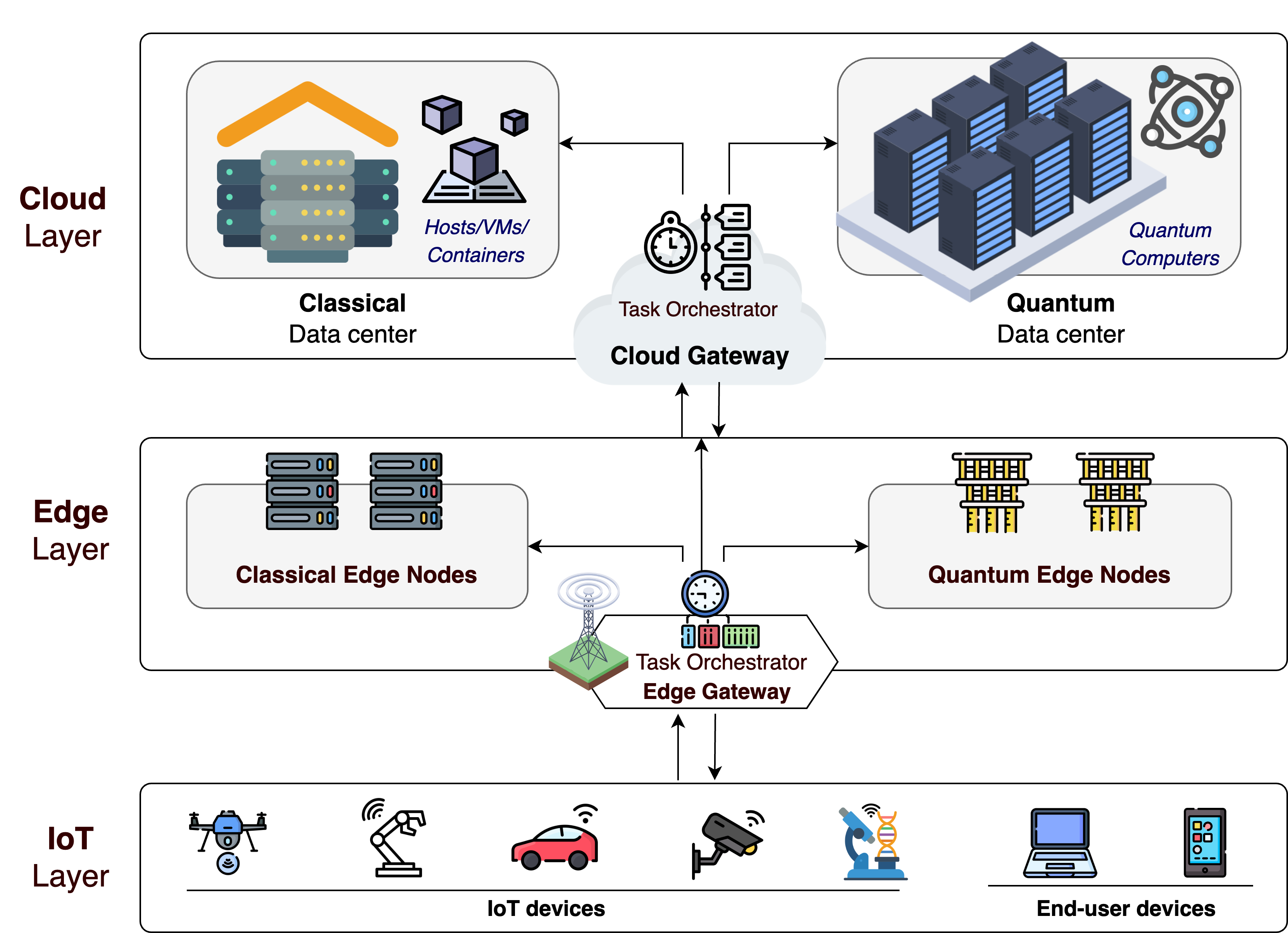}}
\caption{Overview of a Hybrid Quantum Computing Paradigm}
\label{fig:q-hybrid}
\end{figure}

In the classical realm, edge computing has arisen to extend cloud computing capabilities to the edge network \cite{porambage_survey_2018} and enable in-situ processing to reduce the network latency for time-sensitive applications \cite{shi_edge_2016}. As quantum computing matures, we anticipate witnessing the future hybrid quantum paradigm, in which the quantum computation node can be placed in the edge layer, which is closer to the data source and Internet-of-Things (IoT) devices (see Figure \ref{fig:q-hybrid}). The realization of this complete paradigm depends on the advances in quantum hardware development and quantum error correction techniques. As recent years witnessed rapid growth and breakthrough in quantum hardware, we can begin to prepare the system prototype, software, and system algorithm for the near future of hybrid cloud-edge quantum computing \cite{ma_hybrid_2022}. 

In the current quantum cloud paradigm, Quantum Computing as a Service (QCaaS) \cite{qcaas} becomes a primary service model that provides access to quantum computation resources for executing quantum jobs. Cloud vendors offer this service model with a pay-per-use pricing model, where users only need to pay for the actual resources they consume on a per-time unit or per-task basis. Given those near-term quantum computers are subject to noise\cite{nisq-preskill}, quantum executions must be run multiple times (or shots) and corrected with error correction techniques to obtain better results. Thus, the QCaaS model also considers the number of shots to determine the total cost of execution, and this characteristic should be included in the quantum service model.

\subsection{Quantum Programs and their Execution in the Cloud}
A quantum program comprises a sequence of primitive instructions that can be executed on a quantum computer \cite{heim_quantum_2020}. There are two types of quantum programs: gate-based quantum circuits and annealing-based quantum programs \cite{hauke_perspectives_2020}. This study considers the gate-based model, which most quantum hardware vendors, such as IBM Quantum, Google, IonQ, and Rigetti, have widely employed. Based on this model, a quantum circuit is built by performing quantum operations and measurements on qubits using different quantum gates \cite{weder_quantum_lifecycle_2022}.

Figure \ref{fig:q-hybrid} also provides an overview of the hybrid cloud-based quantum-classical computing system. As a quantum program cannot be deployed permanently on a quantum computer for multiple executions \cite{q-api-gateway}, the classical cloud backend is still required in this hybrid system to receive all user requests. Then, a job orchestrator classifies these tasks and forwards them to either classical or quantum computation nodes for execution. We refer to a quantum task that is sent to a cloud-based quantum computer for execution as a \textit{``qulet"}, analogous to a classical task or \textit{``cloudlet"} \cite{calheiros_cloudsim_2011}. While isolated environments such as virtual machines (VMs) or containers can be created within classical servers, there are currently no equivalent techniques for creating quantum VMs or containers. Therefore, qulets must be sent to execute directly on quantum computers each time they are compiled from classical computers.

\subsection{Quantum Computer Performance and Benchmarking}
According to Wack \textit{et al.} \cite{wack_quality_2021}, three metrics are suggested for evaluating the performance of today's quantum computers: the number of qubits, quantum volume (QV), and circuit layer operation per second (CLOPS). The number of qubits determines the scale of a quantum system, \textit{i.e.,} the amount of quantum information that can be processed. In the current era of noisy intermediate-scale quantum (NISQ) devices \cite{nisq-preskill}, the number of qubits is relatively limited, ranging from tens to several hundreds of qubits. For example, at the current time of this study, the IBM Quantum Osprey system had 433 qubits\footnote{https://spectrum.ieee.org/ibm-quantum-computer-osprey}, QuEra's quantum computer had 256 qubits\footnote{https://www.quera.com/aquila}, and IonQ had an 11-fully connected-qubit quantum computer\footnote{https://ionq.com/quantum-systems/harmony}. Apart from the number of qubits, QV and CLOPS are widely used by quantum hardware vendors, such as IBM Quantum and Rigetti, to benchmark the performance of gate-based quantum systems. Quantum volume (QV) is a metric to measure the quality of qubits of a quantum system. Cross \textit{et al.} \cite{cross_validating_2019} defined the formula of quantum volume as $V_Q = 2^{\min(d,m)}$ where $d$ and $m$ are the depth and width of the largest square circuit ($d=m$) that can be precisely executed. Besides, CLOPS is the most recent proposed metric to measure the speed of a quantum system by determining how many QV circuits can be successfully executed per second \cite{wack_quality_2021}. CLOPS is empirically measured by $(M \times K \times S \times D)/time\_taken$ where  $M=100, K=10, S=100, D = \log_2{QV}$ which stands for the number of templates, number of parameter updates, number of shots, and number of QV layers, respectively. Similar to FLOPS (floating point operations per second) or MIPS (million instructions per second) in classical computing, CLOPS is a potential metric that can be used to estimate the running time of a quantum job in a specific system.

\section{Modeling Quantum Computing Environments}
\label{sec:modeling-qcloud}
\begin{figure*}[htbp]
\centerline{\includegraphics[scale=0.085]{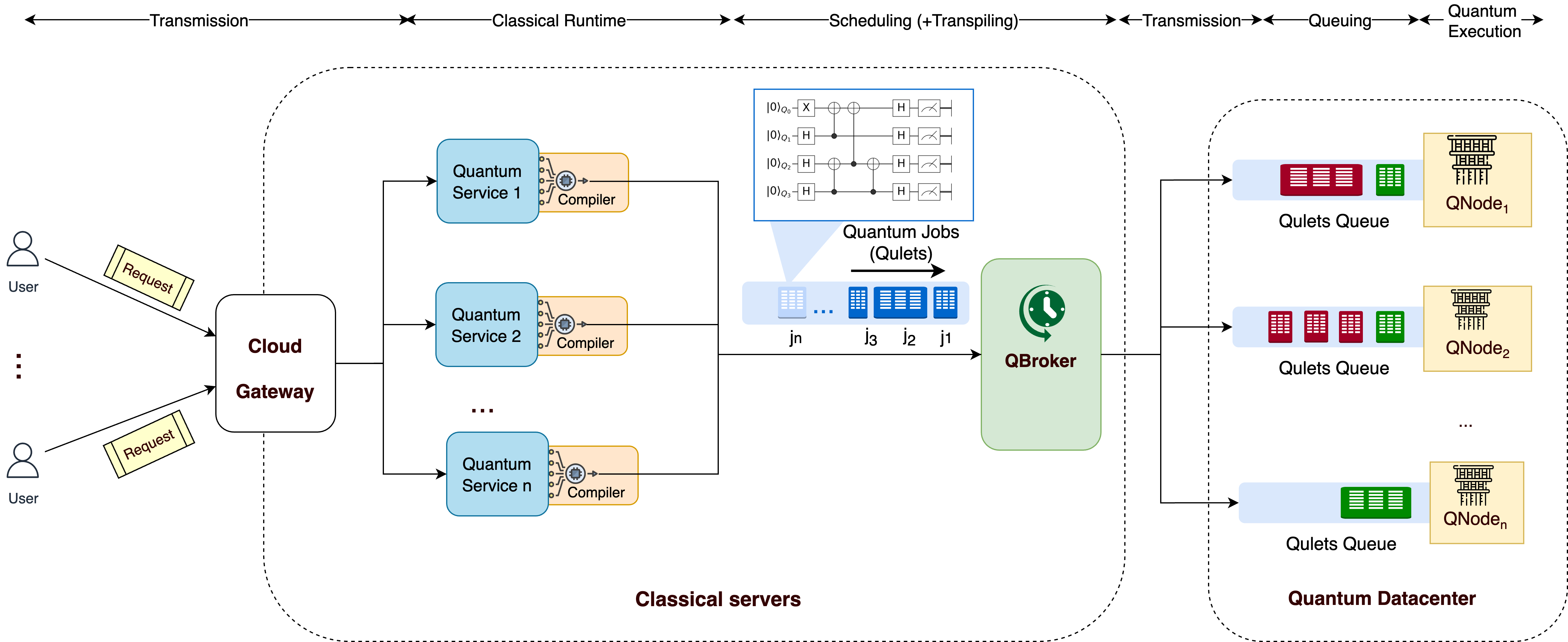}}
\caption{Overview of the System model in Hybrid Quantum Computing Environments}
\label{fig:qcloud-overall}
\end{figure*}

As cloud-based quantum computing is still an emerging paradigm of distributed computing, there is a lack of established standards, models, and methodologies that can proficiently effectively optimize resource management while also dealing with the intricate infrastructure and application challenges. In this section, we propose the first system model for quantum computing environments based on the characteristics of the state-of-the-art quantum computing paradigm. This system model facilitates the design and development of iQuantum and research in quantum distributed systems.

Figure \ref{fig:qcloud-overall} illustrates the overall quantum cloud service model, including the main components of the quantum cloud computing paradigm. The core components of the quantum cloud environment can be modeled as follows:
\begin{enumerate}    
    \item \textit{Quantum Data center} ($\mathcal{D}_\mathcal{Q}$) is a centralized hub that manages a set of quantum systems which are accessible through a classical cloud or edge computing system. The representation of a quantum datacenter is as follows: $\mathcal{D}_\mathcal{Q} = {\mathcal{Q}_1, \mathcal{Q}_2,..., \mathcal{Q}_n}$, where $\mathcal{Q}_i$ represent a quantum computation node in the data center. These quantum nodes can be either homogeneous or heterogeneous based on physical quantum nodes' characteristics and underlying technology. Most quantum cloud providers also provide access to their quantum simulators. However, as they are classical-based resources that facilitate the testing phase in the NISQ era but not the production phases for the long term, we do not consider these simulators in the quantum computing model.
    
    \item \textit{Quantum Computation nodes} ($\mathcal{Q}$) are physical quantum computers located in a specific quantum data center and take the key responsibilities for executing all quantum tasks.  A quantum system can be modeled by a set of properties $\mathcal{Q}_i = \{q^w,q^v, q^s, q^g, q^t, q^e\}$, where:
    \begin{itemize}
        \item $q^w$ is the number of qubits, which indicates the scale of the quantum system.
        \item $q^v$ is the quantum volume of the system, which measures the quality of qubits and the system's capability to execute a quantum circuit faithfully.
        \item $q^s$ refers to the quantum computation speed of the system, which is measured by using CLOPS metric. The higher CLOPS, the faster the system can perform quantum computations and the more complex problems it can tackle.
        \item $q^g$ is the list of the quantum gate sets supported by the quantum system.
        \item $q^t$ is the qubit topology, which includes a list of all connections between pairs of qubits within the quantum chip. The qubit topology describes how the qubits are arranged and interconnected and is an important factor in determining the quantum chip's capabilities.
        \item $q^e$ indicates the quantum system's error rates (such as readout and CNOT gate error). This parameter can help to demonstrate a complete picture of NISQ device performance. Indeed, NISQ devices are characterized by their limited number of qubits and high error rates, which can particularly affect the reliability of quantum computations. 
    \end{itemize}
    Besides, unlike classical computing, quantum computing does not currently employ virtualization or containerization techniques. Instead, quantum computers directly execute quantum tasks without delegating to any virtual machines (VMs) or containers that might be allocated inside. As a result, there is no analogous conceptual model for VMs or containers in the quantum computing domain at this time. 
    
    \item \textit{Quantum Tasks (or qulet $\gamma$)} represent units of quantum computation that can be executed on a quantum computation node. The term \textit{qulet} is analogous to \textit{cloudlet} \cite{calheiros_cloudsim_2011}, which refers to a cloud-based task in classical computing. A qulet is defined by a set of attributes 
    $\gamma_i=\{\gamma^a, \gamma^g, \gamma^w, \gamma^d, \gamma^s, \gamma^e, \gamma^t\}$, where:
    \begin{itemize}
        \item $\gamma^a$ is the submission time of qulet
        \item $\gamma^g$ is quantum gate set in the circuit, $j^g = G_s \cup G_m$ where $G_s$ is a list of single-qubit gates and $G_m$ is a list of multiple-qubit gates.
        \item $\gamma^w$ is the number of qubits (circuit width).
        \item $\gamma^d$ is the number of circuit layers (circuit depth).
        \item $\gamma^s$ is the number of shots (\textit{i.e.,} execution repetition).
        \item $\gamma^t$ is the qubit topology in the circuit.
    \end{itemize}
    If the error rates are considered, each quantum task can be accommodated with the Quality-of-Service (QoS) metrics, which indicates the acceptable error threshold for its execution in the NISQ device.
    
    \item \textit{Quantum Computation Broker} (QBroker) is an intermediary entity between the cloud/edge servers and the quantum data center. Its primary responsibility is to schedule qulets to the most appropriate quantum computation nodes based on the qulets' properties and resource availability. This component's design and implementation of scheduling policies can be customized according to specific needs. By effectively scheduling qulets with the most suitable quantum computation nodes, the broker can minimize wait times and maximize resource utilization, thereby improving the overall performance and efficiency of quantum computing. 
\end{enumerate}

\section{iQuantum Design and Implementation}
\label{sec:model-design}
This section describes the design and proof-of-concept implementation of core components in iQuantum to validate the proposed simulator with a sample simulation process.

\subsection{Architectural Design of iQuantum}
\begin{figure}[htbp]
\centerline{\includegraphics[scale=0.08]{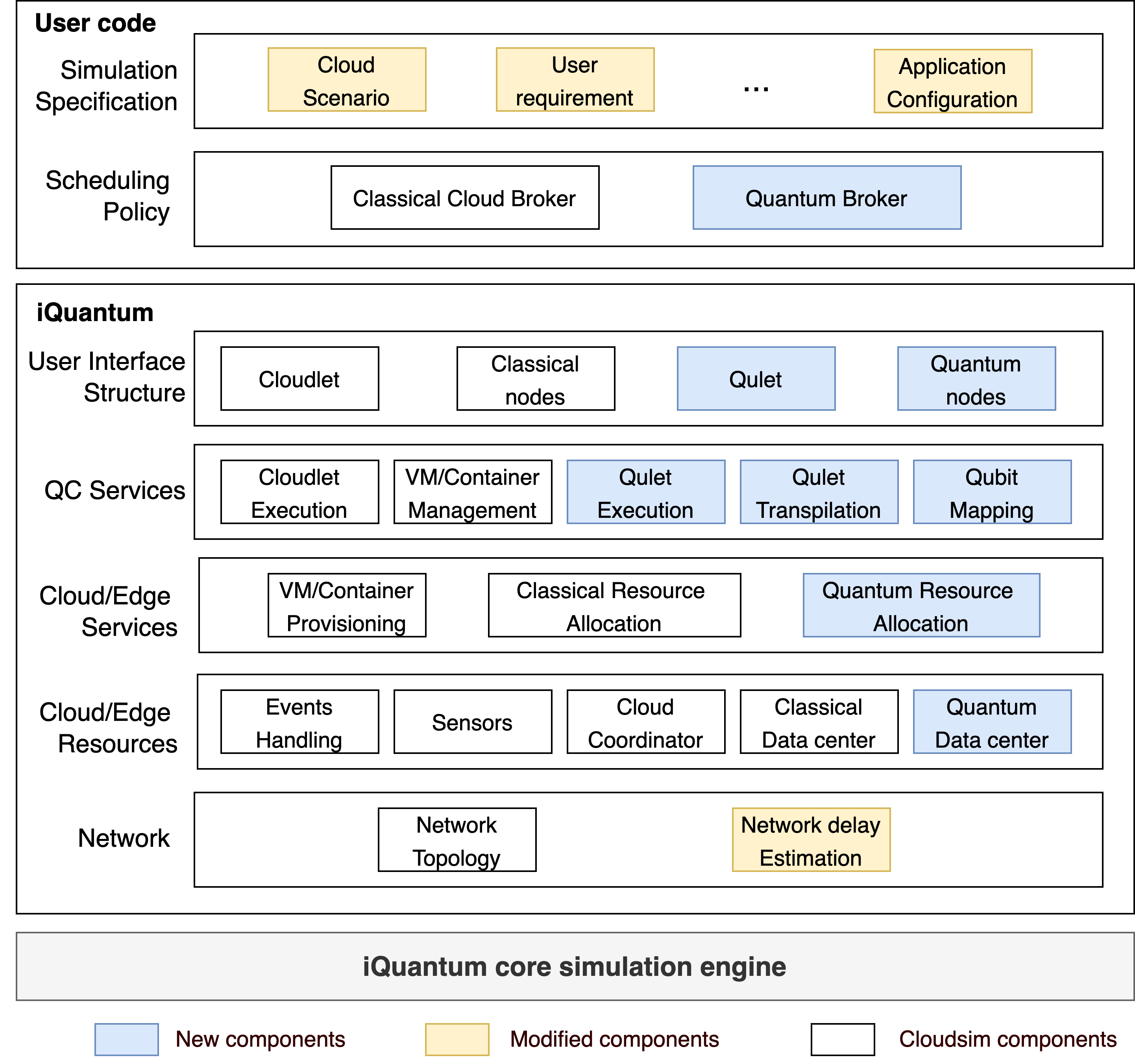}}
\caption{Layered Design of iQuantum}
\label{fig:qcloud-design}
\end{figure}

We proposed the layered design of iQuantum based on CloudSim \cite{calheiros_cloudsim_2011}, which is the open-source simulation framework for system modeling, designing, and evaluating scheduling algorithms for cloud computing environments. The extensions to iQuantum focus on adding quantum features to simulate the hybrid quantum computing environment (as the system model defined in Section \ref{sec:modeling-qcloud}). The layered architecture design of iQuantum is illustrated in Figure \ref{fig:qcloud-design}.  

The User Code layers enable users to define the specific scenarios, requirements, application configurations, and scheduling policies for classical and quantum jobs. Given the unique characteristics of quantum jobs (qulets), which differ significantly from classical jobs (cloudlets), new algorithms with specific constraints are needed to place qulets optimally on suitable quantum nodes. These algorithms should be designed and implemented in the quantum cloud broker. 

The iQuantum simulation layers provide dedicated interfaces for modeling and simulating all core components, as defined in section \ref{sec:modeling-qcloud}, including quantum data centers, quantum nodes, qulets, cloud services, and other resource allocation policies. The specific processes of quantum computing, such as qubit mapping and circuit transpiling, can affect the qulet execution procedure, and strategies to address these challenges can be defined in corresponding classes of iQuantum. Since quantum resources cannot be easily divided into smaller isolated parts, such as virtual machines or containers in classical computing, different resource allocation strategies must be designed to distribute resource usage among multiple users. Furthermore, quantum job execution is expected to incur more network delay, as intermediate transmissions from the classical cloud to the quantum data center are required. Therefore, network delay estimation will be modified to include all encountered delays.

\subsection{Core Implementation of iQuantum}
\label{sec:core-implementation}
\begin{figure}[htbp]
\centerline{\includegraphics[scale=0.11]{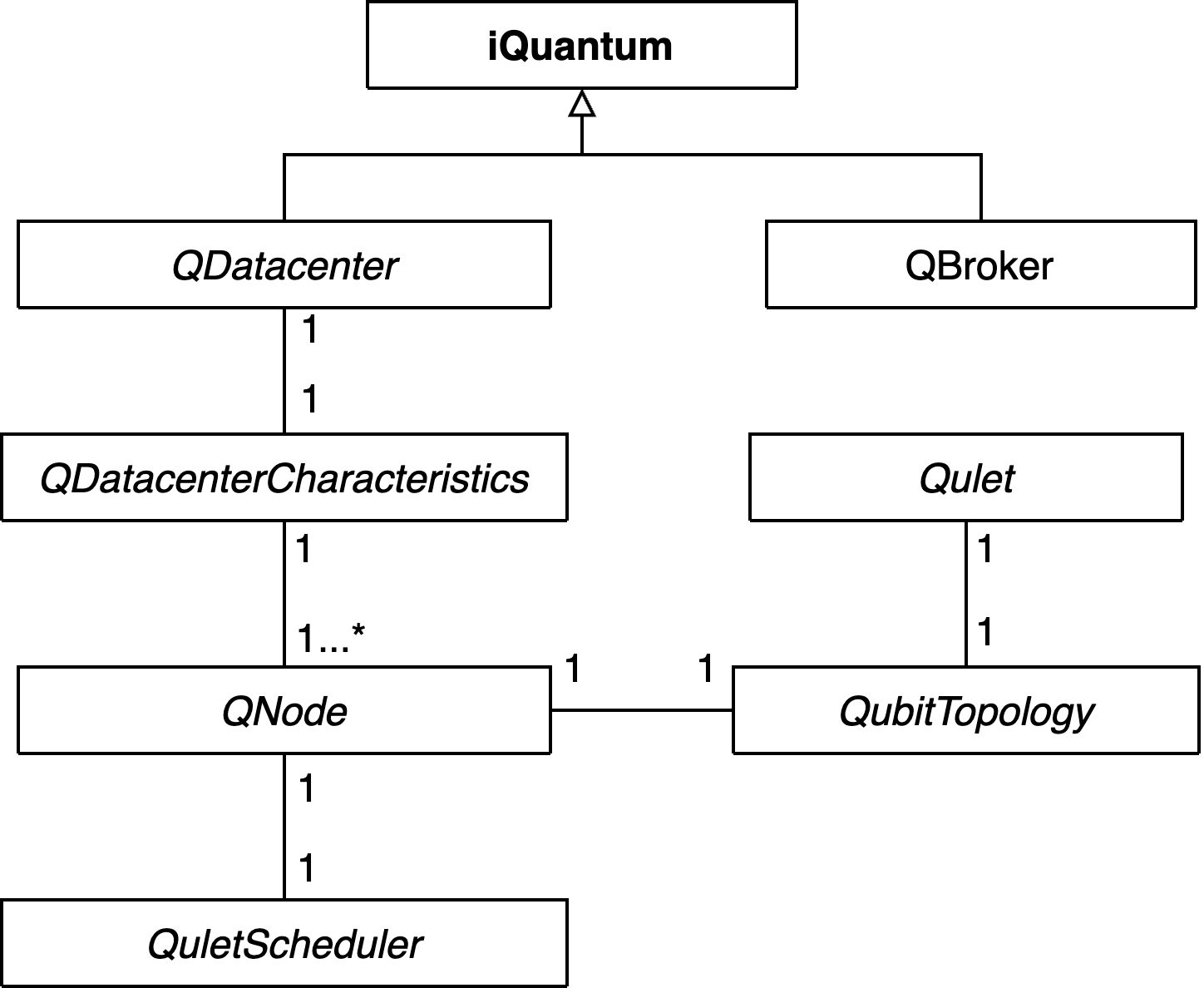}}
\caption{Overview of iQuantum Main Class Diagram}
\label{fig:qcloud-class}
\end{figure}

To validate the proposed simulator, we implement the core components of iQuantum in Java based on the discrete-event simulation (SimEvent) technique of CloudSim \cite{calheiros_cloudsim_2011}. Figure \ref{fig:qcloud-class} shows the overview of the main class diagram of our initial implementation with respect to the proposed quantum computing environment model (Section \ref{sec:modeling-qcloud}). 

\begin{enumerate}
    \item \code{QDatacenter} class models the core infrastructure of a quantum data center, which consists of a collection of quantum nodes (\code{QNode}) used for the qulet execution. All configurations of the quantum data center, including the list of quantum nodes (\code{qNodeList}), are defined using the \code{QDatacenterCharacteristics} class.
    \item \code{QDatacenterCharacteristic} class models the configurations and information of a quantum data center.
    \item \code{QBroker} class models a broker that handles interactions between other components and the core simulator.
    \item \code{QNode} class models a physical gate-based quantum computation instance. It incorporates important metrics such as the number of qubits, quantum volume, CLOPS, gate set, qubit topology, and a scheduling policy.
    \item \code{QubitTopology} class models the connectivity among all qubits of a quantum node or a quantum circuit in qulets. These topologies can be used to design a qubit mapping strategy for qulets to be mapped into a quantum node (discussed in Section \ref{sec:qubit-mapping}).
    \item \code{Qulet} class models a quantum task (or quantum circuit) to be sent to the \code{QBroker} for scheduling and execution in the appropriate \code{QNode}, following the scheduling policy defined in the \code{QuletScheduler} class.
    \item \code{QuletScheduler} is an abstract class implemented by a \code{QNode} instance to model the scheduling policy for determining the share of quantum computation resources among multiple qulets. In the initial implementation, we design the \code{QuletSchedulerSpaceShared} provisioning policy, which allows only one qulet to run in a quantum node at a time, and it must finish execution before scheduling another qulet to the same node. This policy reflects the current situation of cloud-based quantum computing services. However, other policies, such as the Time Shared policy, can be implemented in futher development to allow multiple qulets to share the same quantum node and execute simultaneously.
    
\end{enumerate}

\section{A Sample Simulation using iQuantum}
\begin{figure}[htbp]
\centerline{\includegraphics[scale=0.08]{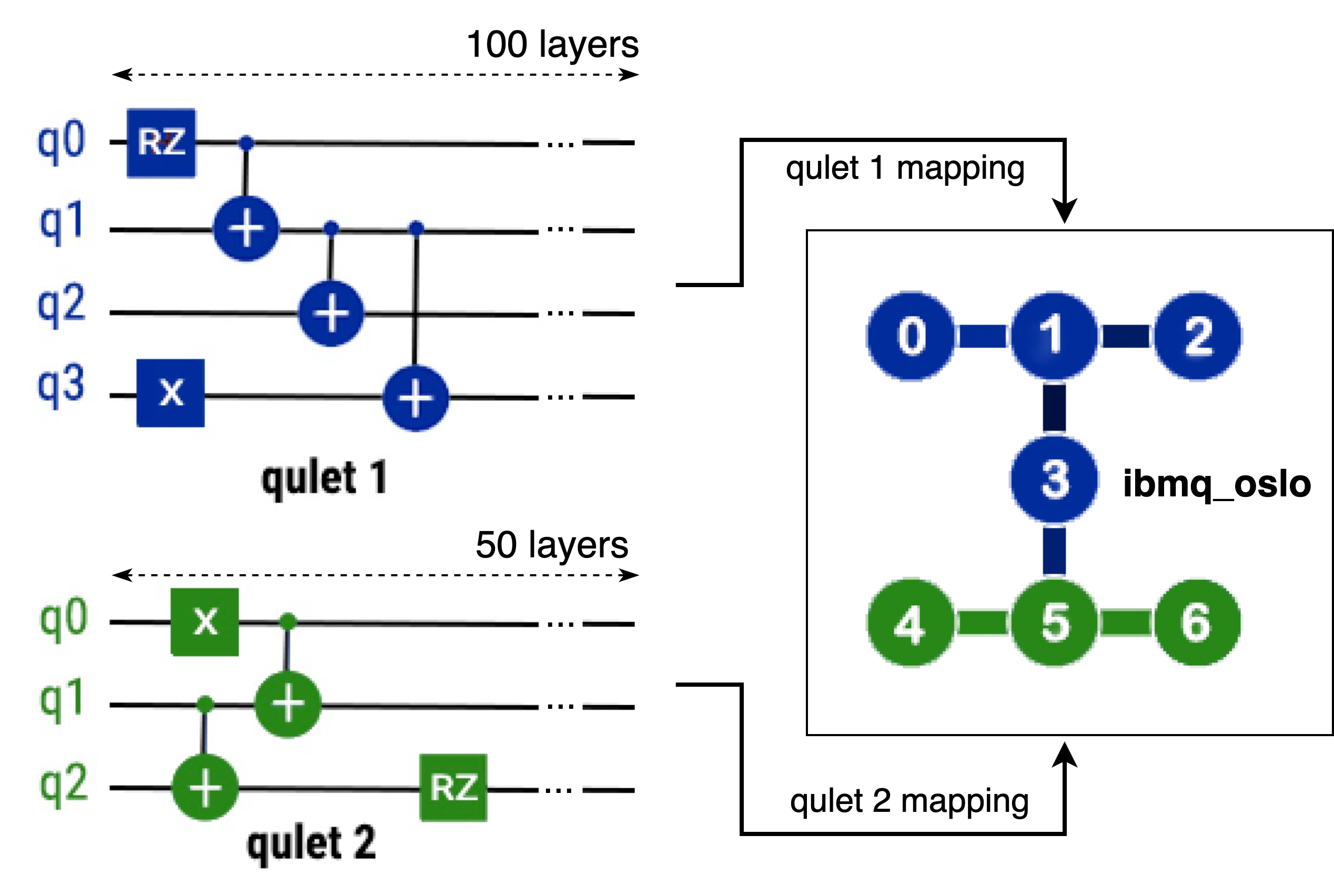}}
\caption{A sample scenario for the simulation (\textit{Left}: a part of quantum circuit in each qulet, \textit{Right}: the qubit topology of the 7-qubit ibmq\_oslo node) }
\label{fig:qubit-mapping}
\end{figure}

This section presents an explanatory example outlining the main steps of executing a simulation scenario in iQuantum. 
Figure \ref{fig:qubit-mapping} illustrates the simple scenario with 2 quantum tasks (qulets) to be submitted to a data center with one quantum node (7 qubits). We model the quantum node using same metrics and qubit topology to IBMQ Oslo node\footnote{https://quantum-computing.ibm.com/services/resources}.    

\textit{Step 1. }Initialize the core simulation instance using \code{iquantum.init()} function.

\textit{Step 2.} Create \code{QNode} instances: Each \code{QNode} instance is associated with different metrics and a qubit topology. For example, the metrics of ibmq\_olso quantum as follows: quantum volume 32, CLOPS 2600, 5 basis gates (CX, ID, RZ, SX, X), and the qubit topology can be represented as $q^t_{oslo} = \{(0,1); (1,2); (1,3); (3,5); (4,5); (5,6)\}$. We models this \code{QNode} in iQuantum as follows (Code \ref{lst:qnode-1}):

    \begin{lstlisting}[language=java, caption=Sample code for modeling a QNode (ibmq\_oslo), label={lst:qnode-1}]
List<int[]> edges_oslo = new ArrayList<>();
edges_oslo.add(new int[]{0, 1});
edges_oslo.add(new int[]{1, 2});
edges_oslo.add(new int[]{1, 3});
... [truncated]
edges_oslo.add(new int[]{5, 6});
QubitTopology osloTpl = new QubitTopology(7, edges_oslo);
ArrayList<String> gateSet = new ArrayList<>(Arrays.asList("CX","ID","RZ","SX","X"));
QNode qNodeOslo = new QNode(7,32,2600, gateSet, osloTpl, new QuletSchedulerSpaceShared());
\end{lstlisting}
   
\textit{Step 3:} Create a \code{QDatacenter} and add \code{QNode} instances to that \code{QDatacenter}. \code{qNodeList} and other information of the quantum datacenter are encapsulated in a \code{QDatacenterCharacteristics} object.  
    \begin{lstlisting}[language=java, caption={Sample code for modeling a QDatacenter with the QNode in step 2}, label={lst:qnode}]
qNodeList = new ArrayList<QNode>();
qNodeList.add(qNodeOslo);
QDatacenterCharacteristics characteristics = new QDatacenterCharacteristics(qNodeList, timeZone, costPerSec);
QDatacenter qDatacenter = new QDatacenter("QDatacenter", characteristics);
\end{lstlisting}

\textit{Step 4:} Create a \code{QBroker} instance and define a list of qulets to be executed (\code{quletList}). The qubit connectivity in the quantum circuits of qulet 1 and qulet 2 are $\gamma^t_1 = \{(0,1); (1,2); (1,3)\}$ and $\gamma^t_2 = \{(0,1); (1,2)\}$, respectively. Both qulets employed 3 basic gates (CX, RZ, X), which is fully supported by the quantum node. Qulet 1 comprises 100 circuit layers and will be executed 4000 shots while these metrics for qulet 2 are 50 layers and 1000 shots. The \code{QBroker} and these qulets can be defined as follows: 
\begin{lstlisting}[language=java, caption=Sample code for modeling a QBroker and a Qulet, label={lst:qnode}]
QBroker qBroker = new QBroker("QBroker");
List<int[]> ql1Edges = new ArrayList<>();
ql1Edges.add(new int[]{0, 1});
ql1Edges.add(new int[]{1, 2});
... [truncated]
QubitTopology ql2Topology = new QubitTopology(3, ql2Edges);
ArrayList<String> qlGates = new ArrayList<>(Arrays.asList("CX", "RZ", "X"));
Qulet qulet1 = new Qulet(0, 5, 100, 4000, qlGates, ql1Topology);
Qulet qulet2 = new Qulet(1, 3, 50, 1000, qlGates, ql2Topology);
\end{lstlisting}

\textit{Step 4:} Design and implement the customized qulet scheduling policy. We implemented a simple \code{QuletSchedulerSpaceShare} policy by extending the \code{QuletScheduler} class. In the initial implementation, we estimated the approximate completion time ($t^q$) of a qulet inside a quantum node by implementing the following equation:
$t^q = \dfrac{\gamma^d}{q^s} \times \gamma^s$ where $\gamma^d$ is the number of circuit layers in the qulet, $q^s$ is the CLOPS of the quantum node, and $\gamma^s$ is the number of shots that qulet need to be executed. However, we acknowledge that other factors, such as transmission time, classical runtime, and queuing time, may also play a role in determining the scheduling policy. We plan to consider these factors and constraints in our future work on job scheduling (discussed in Section \ref{sec:job-scheduling}).

\textit{Step 5:} Submit \code{quletList} to \code{qBroker} and start the simulation. Once all the simulation tasks are complete, stop the simulation and print out the final result.
\begin{lstlisting}[language=java, caption=Sample code for starting the simulation and print out the result, label={lst:qnode}]
qBroker.submitQuletList(quletList);
iquantum.startSimulation();
...[truncated]
iquantum.stopSimulation();
List<Qulet> r = qBroker.getQuletReceivedList();
printQuletList(r);
\end{lstlisting}
The simulator shows all events (with timestamps) happening during the simulation period. 
\begin{lstlisting}[language=java, title={Sample events in the simulation} , label={lst:qnode}]
0.0: QBroker: Cloud Resource List received with 1 resource(s)
0.01: QBroker : Started scheduling all Qulets to QDatacenter 
0.01: QBroker: Sending Qulet 0 to QNode #0
0.01: QBroker: Sending Qulet 1 to QNode #0
153.86: QBroker: Qulet 0 result received
173.09: QBroker: Qulet 1 result received
173.09: QBroker: All Qulets executed. Finishing
\end{lstlisting}
As shown in the simulation results, two qulets are submitted to the QNode at timestamp t = 0.01s (minimum interval between 2 events). The execution time of qulet 1 and qulet 2 in the \code{QNode} are 153.85s and 19.23s, respectively. According to the Space Share scheduling policy, qulet 2 is executed after qulet 1 finish its execution. Therefore, the total execution time of all qulets is 153.85 + 19.23 = 173.08s. More complex scenario with advanced scheduling policies and qubit mapping techniques (discussed in Section \ref{sec:qubit-mapping}) can be implemented to allow concurrent execution of multiple qulets, optimize the total execution time and the resource utilization. 

\section{Potential Use Cases and Extensions}
This section outlines potential applications and further extensions of iQuantum in modeling different quantum computing research problems, including quantum job scheduling, qubit mapping, and hybrid quantum-classical system. 

\subsection{Model and Design Quantum Job Scheduling Algorithms}
\label{sec:job-scheduling}
Job scheduling (or task placement) is a critical aspect of distributed system research, including quantum computing, as it helps optimize resource utilization and minimize total job completion time. In a cloud-based quantum system, the responsibility of quantum job scheduling lies with the QBroker, which is tasked with finding the most suitable quantum system to execute a given qulet. To facilitate quantum job scheduling design, users can leverage iQuantum to generate cloud workloads and design and evaluate adaptable job scheduling algorithms for quantum cloud environments. By utilizing iQuantum, researchers can simulate various quantum computing scenarios and evaluate the effectiveness of different job scheduling strategies, helping to improve the performance and efficiency of the quantum computing system.

\begin{figure}[htbp]
\centerline{\includegraphics[scale=0.10]{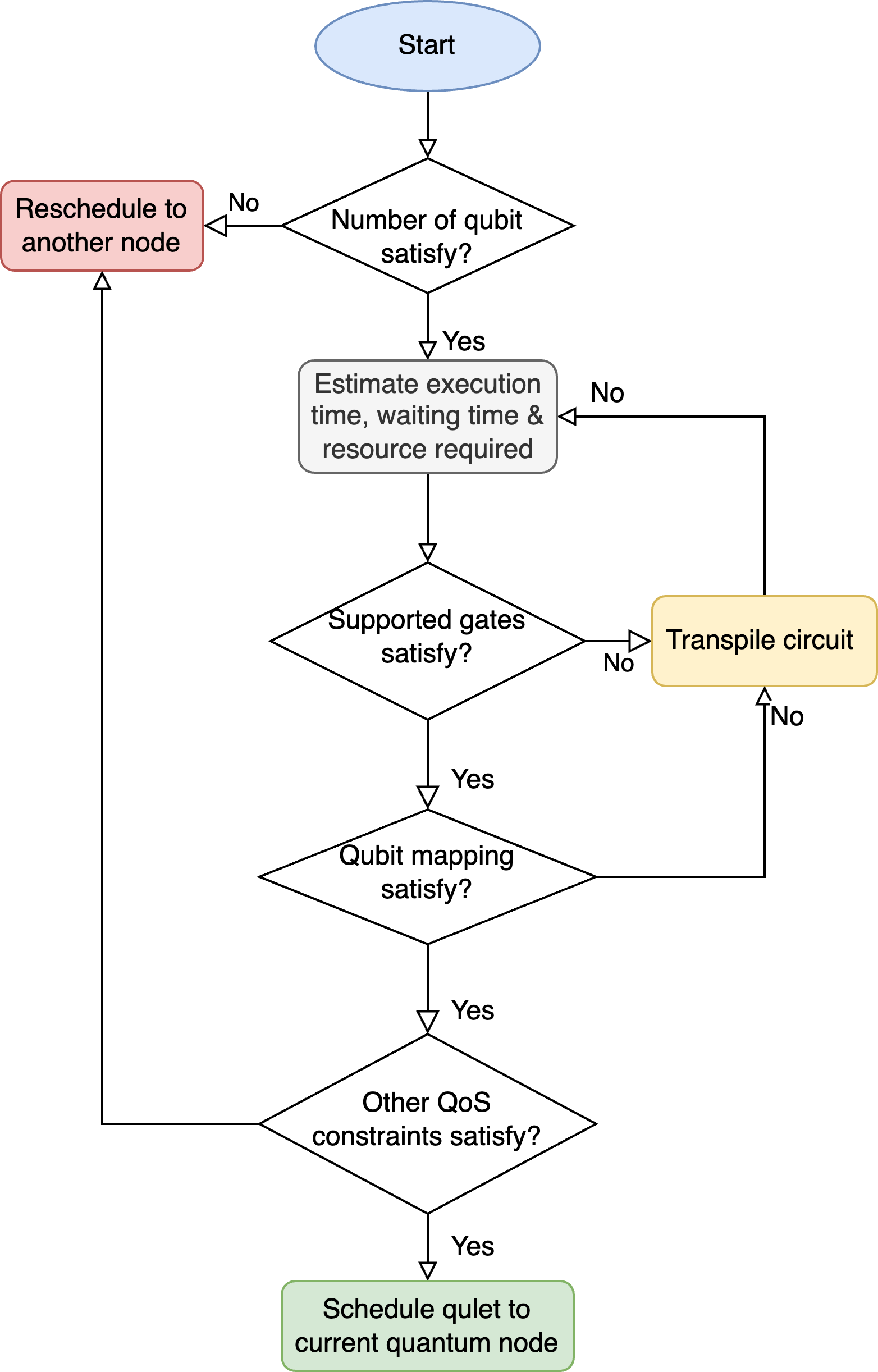}}
\caption{Sample qulet scheduling logic with multiple constraints}
\label{fig:job-scheduling}
\end{figure}

A placement configuration of job $\gamma_i \in \Gamma$ can be define as $\sigma_i = \{\gamma_i, q_k\}$ where $q_k \in \cal{Q}$ and $1 \le k \le |\cal{Q}|$ is the quantum computation node index. Several constraints must be considered when designing quantum job scheduling for specific situations. For example, requirements to place qulet $\gamma_i$ to a target quantum computer $q_T$ as follows (see Figure \ref{fig:job-scheduling}):

\begin{enumerate}
    \item The number of qubits of $q_T$ must be greater or equal to the number of the required qubits to execute qulet $\gamma_i$: $q^w_T \ge \gamma_i^w$. This requirement must be satisfied when selecting a quantum computation node to execute a qulet. If the backend does not meet this requirement, the qulet must either be rescheduled to another quantum computation node or divided into multiple smaller circuits using techniques such as circuit cutting \cite{tang_cutqc_2021} to fit the capacity of the targeted quantum system.
    
    \item To execute qulet $\gamma_i$ on quantum computation node $q_T$, all gates used in $\gamma_i$ must be supported by $q_T$, \textit{i.e.,} $\gamma_i^g \in q_T^g$. If an unsupported gate is detected, it must be decomposed into multiple native gates of the target machine (using circuit transpilation technique \cite{aravanis_transpiling_2022}). However, this approach may increase the cost of execution since the processing time may be longer.
    
    \item The qubit connectivity in the circuit of $\gamma_i$ must be a subset of the qubit topology of the target quantum system $q_T$, \textit{i.e.,} $\gamma_i^t \subset q_T^t$. If the connectivity does not meet this requirement, the input circuit must be transpiled and swapped to match the qubit order of the target quantum system. This process is known as qubit mapping and is described in more detail in Section \ref{sec:qubit-mapping}.

    \item Other quality of service (QoS) constraints must also be considered. For example, the estimated execution time $t_i$ of qulet $\gamma_i$ must be less than or equal to the expected completion time declared by users, \textit{i.e.,} $t_{\gamma_i} \le \gamma^e_i$. Another constraint is to ensure that the quantum volume of the target quantum system $q_T$ is large enough to execute the qulet faithfully and accurately.
\end{enumerate} 

To estimate the total execution time $t_{\gamma_i}$ of a quantum job $\gamma_i$, multiple factors need to be assessed (as shown in Figure \ref{fig:qcloud-overall}):
$t_{\gamma_i} = t^n_{\gamma_i} + t^c_{\gamma_i} + t^s_{\gamma_i} + t^w_{\gamma_i} + t^q_{\gamma_i}$, where $t^n$ is transmission times (or network delay), $t^c$ is classical runtime to compile the quantum circuit, $t^s$ is scheduling time (including transpilation time if necessary), $t^w$ is the queuing time at the targeted quantum system, and $t^q$ is the actual quantum execution time inside the quantum system. 

The key objective of quantum job scheduling is to minimize the total execution time of all incoming jobs while optimizing the resource utilization of the whole system. As quantum computing is an emerging paradigm, job scheduling and resource allocation remain active research areas. Several studies have proposed strategies to tackle this issue. For example, Ravi \textit{et al.} \cite{ravi_adaptive_2021} proposed a statistical-based strategy by analyzing quantum job data over two years. Ngoenriang \textit{et al.} \cite{ngoenriang_optimal_2022} and Kaewpuang \textit{et al.} \cite{kaewpuang_stochastic_2022} proposed stochastic resource allocation for distributed quantum computing under multiple uncertainties of circuit characteristics. Commercial platforms such as IBM Quantum currently use a fair-share algorithm to ensure fairness in using quantum resources for the general public\footnote{https://quantum-computing.ibm.com/lab/docs/iql/manage/systems/queue/}. When the quantum computing model becomes more well-defined, and with the support of simulators like iQuantum, we expect to see more research endeavors in academia and industry to design more efficient job scheduling and resource allocation for quantum computing in the future.

\subsection{Model and Design Qubit Mapping Strategies}
\label{sec:qubit-mapping}
\begin{figure}[htbp]
\centerline{\includegraphics[scale=0.13]{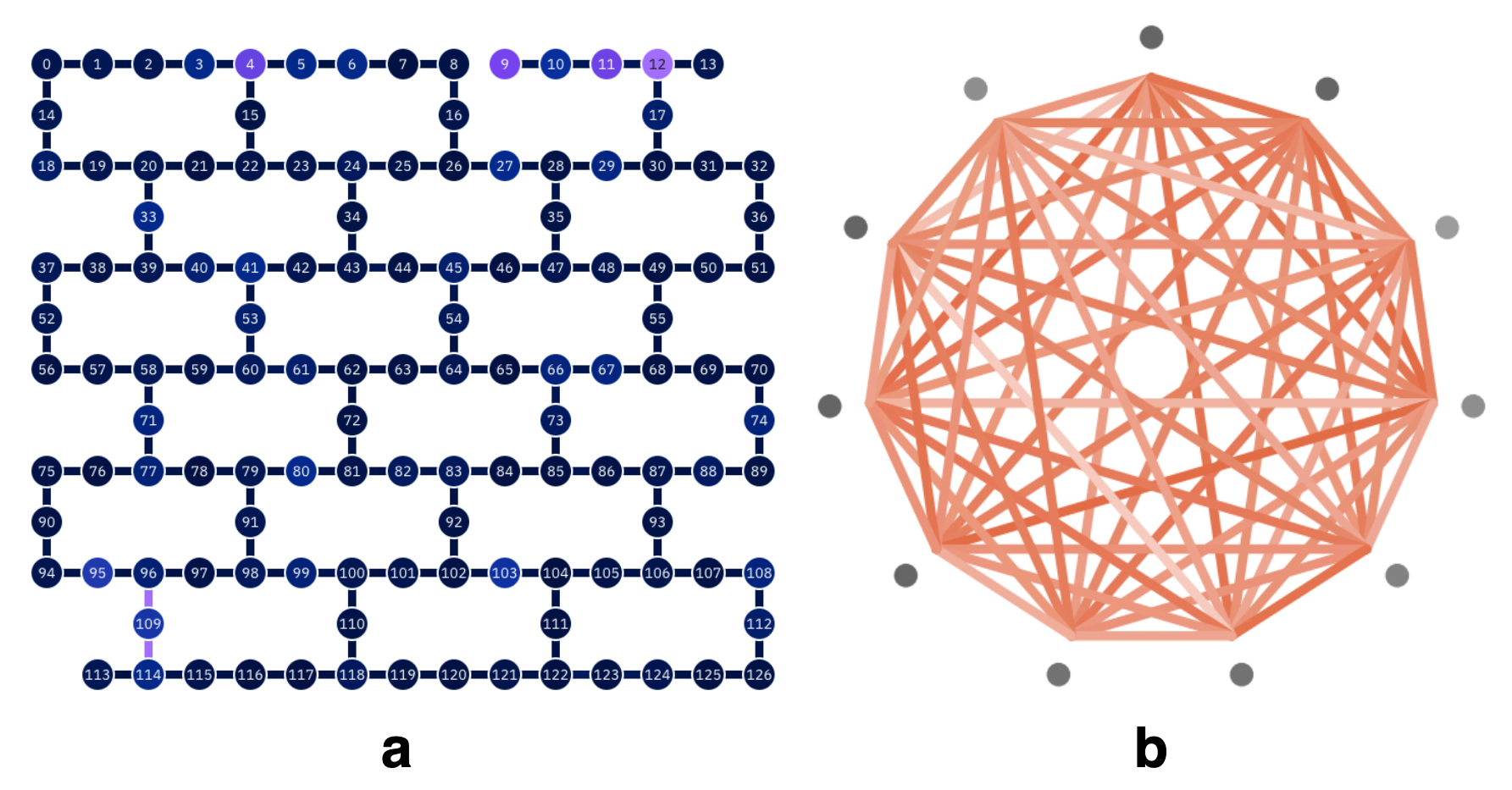}}
\caption{Qubit Topology of (a) 127-qubit IBM Quantum Eagle (ibm\_washington) and (b) 11-qubit IonQ's quantum computer \cite{wright_benchmarking_2019}.}
\label{fig:qubit-ibmq}
\end{figure}

Qubit mapping (or quantum circuit mapping) is a mechanism that ensures the compatibility of quantum circuits with the target quantum computer \cite{liu_qucloud_2022}. As discussed earlier, this mechanism also plays an important role in quantum job scheduling. Each gate-based quantum computer can have a different qubit topology, which refers to the connectivity of all qubits due to different system designs and technologies used. For example, IBM Quantum Eagle chips have 127 partially connected qubits, while IonQ's quantum computer has 11 fully-connected qubits \cite{wright_benchmarking_2019} (see Figure \ref{fig:qubit-ibmq}). A quantum circuit can only be executed if its qubit connectivity can be mapped to the qubit topology of the targeted quantum chip and if the target quantum system supports all single-qubit and multi-qubit gates in the circuit. Otherwise, circuit transpilation techniques can be used to transform the given circuit to match the qubit topology and gate set supported by the target quantum device \cite{younis_quantum_2022}. Although qubit mapping becomes a challenging task when the number of qubits increases, it is an essential approach for optimizing the utilization of quantum cloud resources and enabling the concurrent processing of multiple quantum circuits in today's noisy quantum systems.

We plan to expand the capabilities of iQuantum by enabling users to design and experiment with the circuit mapping technique. With this feature, users can explore various mapping strategies and evaluate their effectiveness in terms of reducing execution time and improving resource utilization on different quantum devices. For example, in our sample simulation scenario (see Figure \ref{fig:qubit-mapping}), two qulets can be allocated simultaneously to \textit{ibmq\_oslo} to maximize the quantum resource utilization of this node and can reduce the total execution time. However, when the number of qubits increases and connectivity among them in both quantum systems and qulets becomes more complicated, the design of the qubit mapping algorithm for concurrent qulets becomes more challenging, as shown in some recent studies, such as \cite{liu_qucloud_2021, liu_qucloud_2022}. Hence, iQuantum can serve as the testbed environment for experimenting with new ideas in qubit mapping techniques.  

\subsection{Model Hybrid Quantum-Classical Computing Environments and Hybrid Task Orchestration Algorithms}
\begin{figure}[htbp]
\centerline{\includegraphics[scale=0.11]{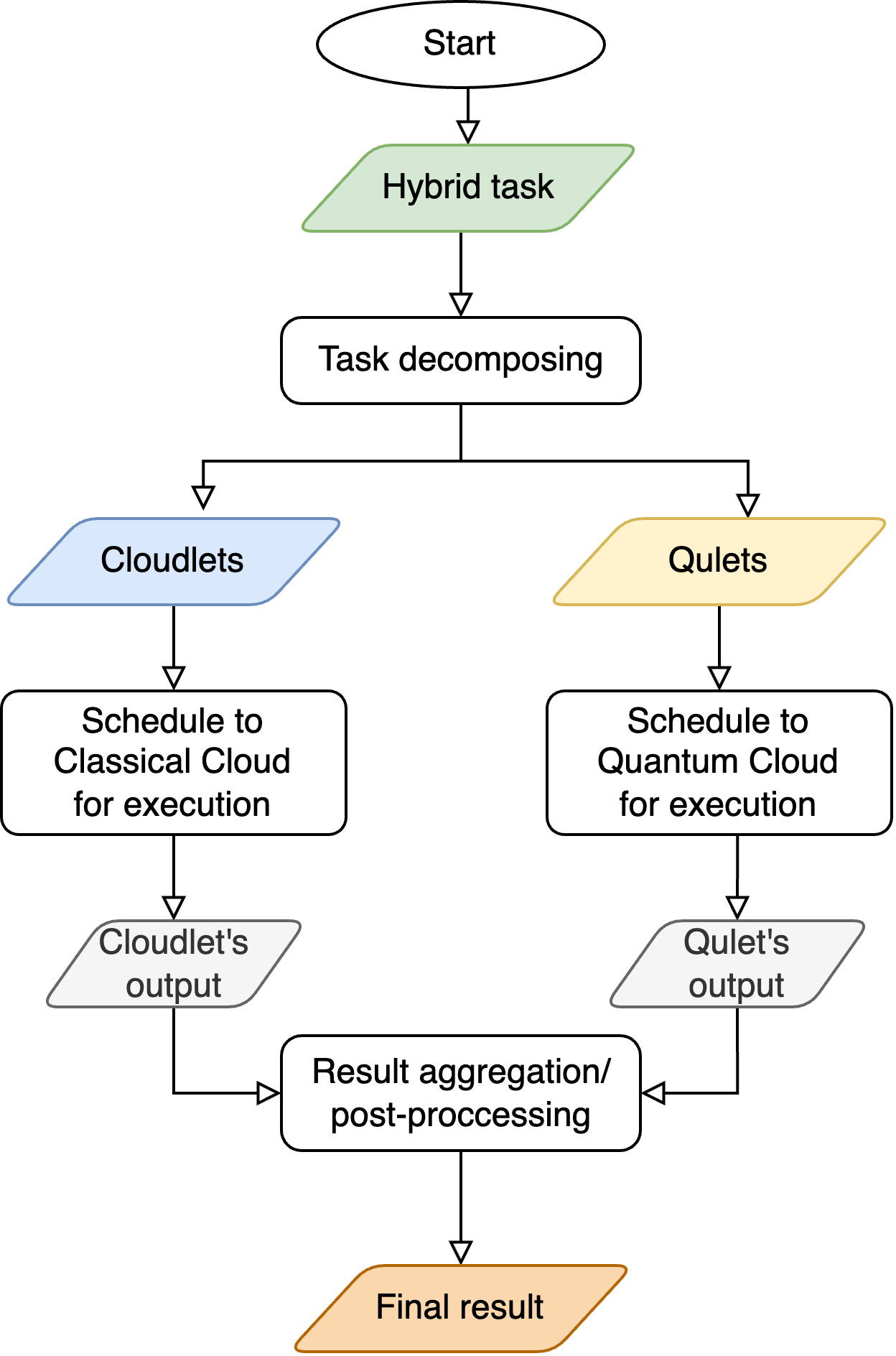}}
\caption{An example of Hybrid Quantum-Classical Task Execution Logic}
\label{fig:hybrid-job}
\end{figure}

Quantum computing is inevitably still in its infancy and cannot replace classical computing systems completely in the near future. Instead, quantum computing is expected to be a complementary technique for classical computing, where each technique will be to execute best-suited tasks to their respective advantages. Thus, the hybrid quantum-classical computing system is a potential approach to leverage the capabilities of both quantum and classical systems in the NISQ era \cite{serrano_quantum_2022}. This paradigm leverages the strengths of classical computing for tasks such as data pre-processing and post-processing while delegating more computationally-intensive tasks to quantum systems \cite{weder_quantum_lifecycle_2022}. The outcomes of these quantum computations can then be combined with classical processing to obtain the final result. Figure \ref{fig:hybrid-job} shows an example of hybrid quantum-classical execution logic. 

Each hybrid task comprises quantum and classical parts, which can be decomposed and distributed to different backends for execution. By building iQuantum on top of CloudSim \cite{calheiros_cloudsim_2011}, we can leverage the existing features of CloudSim to model classical cloud components, such as data centers, hosts, virtual machines, containers, and cloudlets (classical tasks). The remaining quantum components can be modeled using iQuantum features and seamlessly integrated with classical ones to form a hybrid system. We can design a hybrid task orchestration technique for scheduling both \textit{cloudlets} and \textit{qulets} to their respective backend to optimize resource utilization and total completion time for the whole system. As these cloudlets and qulets can be either independent or dependent on each other based on the application design, each task may need to wait for its dependent task to be complete before performing final aggregation or post-processing all outputs and generating the final result. iQuantum is expected to provide a simulation environment for modeling hybrid quantum-classical systems and testing new hybrid task orchestration for these systems.

\section{Related Work}
Table \ref{tab1} summarizes the overall comparison of our proposal and other related work. As far as we know, there is no work in the literature that can be used to model systems and tasks in hybrid quantum computing environments for designing scheduling and resource orchestration algorithms like our proposed simulator in this paper. 

\begin{table}[htbp]
\caption{A feature comparison overview of related works with ours}
\begin{center}
\begin{tabular}{|p{0.1\textwidth}|p{0.07\textwidth}|c|c|c|}
\hline
\textbf{}& \textbf{Simulation} &\multicolumn{3}{|c|}{\textbf{Systems and Tasks Modeling}} \\
\cline{3-5} 
\textbf{Simulators}& \textbf{ Focus} & \textbf{\textit{Quantum}}& \textbf{\textit{Cloud}}& \textbf{\textit{Fog/Edge}} \\
\hline
QuNetSim \cite{diadamo_qunetsim_2021} &  Quantum Network & E & $\times$ & $\times$ \\ \hline
NetSquid \cite{coopmans_netsquid_2021} &  Quantum Network & E & $\times$ & $\times$ \\ \hline
QuEST \cite{jones_quest_2019} & Quantum Operation & E & $\times$ & $\times$ \\ \hline
QXTools \cite{brennan_qxtools_2022} & Quantum Operation & E  & $\times$ & $\times$ \\  \hline
PAS \cite{bian_pas_2021} & Quantum Operation & E & $\times$ & $\times$ \\ \hline
\begin{tabular}[]{@{}l@{}}iQuantum\\ \textit{(Our proposal)}\end{tabular}  & System Modeling & S & \checkmark & \checkmark \\

\hline
\multicolumn{5}{l}{\checkmark: Supported; $\times$: Unsupported; E: Emulation; S: Simulation} \\
\multicolumn{5}{l}{(Cloud/edge features derived from CloudSim and iFogSim)}
\end{tabular}
\label{tab1}
\end{center}
\end{table}

In the classical domain, CloudSim \cite{calheiros_cloudsim_2011} is a popular simulator for modeling cloud computing environments. Similarly, iFogSim \cite{mahmud_ifogsim2_2022} is a well-known simulator for modeling edge and fog computing environments. However, these simulators do not support modeling quantum computing environments, as hybrid quantum computing involves the collaboration of quantum and classical resources. Therefore, this proposed iQuantum will help overcome the limitations of current modeling simulators.

Regarding quantum computing environments, several studies have focused on quantum network protocol simulation. Diadamo \textit{et al.} \cite{diadamo_qunetsim_2021} proposed QuNetSim as a framework for simulating different quantum network protocols, such as quantum key distribution and quantum routing. As QuNetSim relies on other qubit simulators (such as SimulaQron \cite{dahlberg_simulaqron_2019}, ProjectQ \cite{steiger_projectq_2018}, and QuTiP \cite{johansson_qutip_2012}), its main objective is to develop quantum network protocol simulation rather than distributed quantum modeling. Similarly, NetSquid \cite{coopmans_netsquid_2021} is a discrete-event-based network simulator for modeling and simulating quantum network protocols. Jones \textit{et al.} proposed QuEST \cite{jones_quest_2019} as another open-source software for simulating the behavior of quantum systems with high performance. QXTools \cite{brennan_qxtools_2022} is another Julia-based framework for simulating distributed quantum circuits using the tensor networking approach. Also, Bian \textit{et al.} \cite{bian_pas_2021} proposed PAS as a lightweight quantum simulator that works like other qubit simulators. For clarification, we categorized these tools as \textit{Emulation} frameworks as they support emulating actual quantum operations, while our simulation framework supports modeling and simulation of quantum computing environments using the discrete-event system simulation technique. Therefore, our new approach for the iQuantum simulator focused on empowering researchers to conduct performance evaluation experiments of system design and resource management algorithms for quantum computing.

\section{Conclusions and Future Work}
In this paper, we presented the proposal of iQuantum as a first-of-its-kind simulator for modeling hybrid quantum computing environments. Besides, we introduced the system model of key elements in quantum computing systems, including quantum data centers, quantum computation nodes, qulets, and quantum brokers. We also presented the proof-of-concept implementation and a simulation example to validate the proposed simulator. Furthermore, we discussed potential use cases and further development of iQuantum in terms of supporting research in algorithm development for quantum job scheduling, qubit mapping, and modeling hybrid quantum-classical environments. 

As quantum computing evolves, more standards, well-known service models, and system algorithms are expected to be gradually established. We will improve our simulator based on the discussed use cases to adapt to new advances in quantum cloud and distributed quantum computing. Other characteristics of near-term quantum devices, such as noises, error correction, quantum network communications, and the pricing model, will also be considered in the future development of iQuantum.

\textit{Software availability:} The iQuantum software with the source code will be published on our website (\href{http://clouds.cis.unimelb.edu.au/iquantum}{clouds.cis.unimelb.edu.au/iquantum}). Sample programs and tutorials illustrating the use of iQuantum will also be available on the project website.



\bibliographystyle{ieeetr}
\bibliography{references}

\end{document}